# Anomalous Hall effect in antiferromagnetic/non-magnetic interfaces

M. Asa,[1] C. Autieri[2,3], R. Pazzocco[1], C. Rinaldi,[1] W. Brzezicki[3], A. Stroppa[4], M. Cuoco[5], S. Picozzi[2] and M. Cantoni[1*]

[1] *Department of Physics, Politecnico di Milano, Via G. Colombo 81, 20133, Milano, Italy*

[2] *Consiglio Nazionale delle Ricerche CNR-SPIN, c/o Univ. "G. D'Annunzio", 66100 Chieti, Italy*

[3] *International Research Centre MagTop, Institute of Physics, Polish Academy of Sciences, Aleja Lotnikòw 32/46, PL-02668 Warsaw, Poland*

[4] *Consiglio Nazionale delle Ricerche CNR-SPIN, c/o Univ. L'Aquila, 67100 L'Aquila, Italy*

[5] *Consiglio Nazionale delle Ricerche CNR-SPIN, c/o Universitá di Salerno- Via Giovanni Paolo II, 132 - 84084 - Fisciano (SA), Italy*

**ABSTRACT**

We report a combined theoretical and experimental investigation of magnetic proximity and Hall transport in Pt/Cr bilayers. Density functional theory indicates that an interfacial magnetization can be induced in the Pt layer and a strong magnetocrystalline anisotropy with an easy axis out of plane arises in the antiferromagnet. A signal ascribed to the anomalous Hall effect is detected and associated to the interface between Pt and Cr layers. We show that this effect originates from the combination of proximity-induced magnetization and a nontrivial topology of the band structure at the interface.

---

[*] Author to whom correspondence should be addressed: matteo.cantoni@polimi.it

**TEXT**

Anomalous Hall Effect (AHE) is typically associated to ferromagnetic metals because of broken time reversal symmetry and spin-orbit coupling (SOC) [1]. Recent observations of large AHE originating from a non-vanishing Berry phase in antiferromagnets with non-collinear spin arrangement [2,3] brought great interest for possible applications in the emerging field of antiferromagnet spintronics [4]. It is however generally accepted that ordinary antiferromagnets with collinear spin configuration should not exhibit AHE due to vanishing Berry phase.

In this work, we show how interfacing a collinear antiferromagnet (AFM) with a non-magnetic (NM) metallic layer can produce a detectable AHE, whose sign is controllable by field-cooling, in the NM/AFM system Pt/Cr. We propose a mechanism for this phenomenon based on Magnetic Proximity Effect (MPE) in Pt [5] and non-trivial band topology at the interface. Chromium is the prototypical itinerant antiferromagnet, characterized by a spin-density-wave (SDW) magnetic ordering, involving a periodic modulation of the amplitude of antiferromagnetically coupled magnetic moments [6]. Platinum is a non-magnetic metal in its bulk form, on the verge of Stoner ferromagnetism. In thin film heterostructures, MPE has been demonstrated by X-ray Magnetic Circular Dichroism (XMCD) in systems where Pt interfaces FM 3d metals [7–10]. MPE has also been invoked to explain some transport properties of Pt on insulating Yttrium Iron Garnet, [11] although this is still debated [12]. More recently, MPE in Pt has been exploited to detect the state of a neighboring magnetoelectric antiferromagnet $Cr_2O_3$ by means of the AHE [13,14], demonstrating the use of Pt as valuable probe in the context of the emerging AFM-based spintronics [4].

By combining ab-initio calculations and electrical measurements, we demonstrate that a spin-polarized interface forms between Cr and Pt. Density functional theory (DFT) predicts a net magnetic moment induced in Pt by neighboring Cr with out of plane anisotropy. We show the electrical detection of AHE also in relatively thick metallic heterostructures, despite the interfacial origin of such effect. Exploiting the peculiar transport properties of Chromium related to SDW antiferromagnetism [15,16], we can unambiguously associate the presence of AHE to the AFM state of Cr. Finally, we discuss how the presence of AHE and its temperature dependence can be justified by the combined effect of interfacial magnetization and band topology, which could be in principle expected also in other NM/AFM systems.

To model the Pt/Cr interface, we performed first-principles density functional calculations within the Gradient Generalized Approximation by using the plane wave VASP [17] package and the PBE for the exchange-correlation functional [18] including SOC. The core and the valence electrons were treated with the projector augmented wave method [19] and a cutoff of 450 eV for the plane wave basis. In supercell calculations we use 8 Cr layers and the lattice constant of Cr bulk (2.89 Å). (See Supplemental Material S.1 for details).We assume that the spin density wave in Cr [20] will not affect the properties of the Pt/Cr interface, due to the small deviation from commensurate antiferromagnetism. In this 3d-5d interface we neglect the possibility of complex magnetism [21,22] since it was shown that the Dzyaloshinskii-Moriya interaction is negligible at the Cr/Pt interface [22].

Bulk Pt presents a van Hove peak close to the Fermi level. In general the interface hybridization and the magnetic proximity tends to favor a Stoner-type ferromagnetic instability [23], so that Pt undergoes a magnetic phase transition at interfaces. We calculate the profile magnetization of the Pt(N)/Cr heterostructure in the collinear approximation as function of the number (N) of Pt layers. The Pt interface always couples ferromagnetically with Cr, as for 3d ferromagnetic materials [24]. From a theoretical point of view, the magnetic proximity demonstrates that the spin polarization, arising into the antiferromagnet, penetrates the Pt layers. Although the spatial profile is non-monotonous, the sum over the Pt layers leads to a non-vanishing total magnetization of the order of 0.3 $\mu_B$ (See Supplemental Material S.2 for the Cr/Pt magnetic profile).

The magnetocrystalline anisotropy (MCA) is calculated from first-principles (including SOC) as a function of the lattice constant. As illustrated in Fig. 1, the MCA shows a strong sensitivity to strain effects, highlighted by the non-monotonic behavior and in agreement with what previously found for 3d ferromagnets interfaced with 5d materials [25]. For the lattice constant of bulk Cr, a large MCA with an out of plane easy axis is obtained. The contribution to the MCA mainly comes from the first layer of Cr, as demonstrated by means of DFT evaluation of density of states (DOS) (See Supplemental Material S.1 for the Cr/Pt DOS).

Epitaxial Pt/Cr bilayers were grown by Molecular Beam Epitaxy on MgO (001) substrates to experimentally verify the presence of AHE. Cr thin films were deposited in ultra-high vacuum conditions with thickness ranging from 25 nm to 75 nm. Details on the growth of the heterostructure will be published elsewhere. Residual strain as measured from X-ray diffraction is less than 0.4% even in the thinnest film, thus largely comparable to the bulk case. According to first-principles calculations, an MCA with out of plane easy axis is thus expected. Finally, a Pt overlayer with thickness between 0.6 (4 monolayers) and 3 nm was deposited on top. Fifty microns wide Greek crosses were then defined on the samples by optical lithography and ion milling. The compensation of geometrical offsets in this geometry has been achieved by subtracting measurements in which current and voltage contacts are exchanged [26]. At the same time, the current reversal method allows to compensate for thermoelectric offsets (See Supplemental Material S.3 for the measurements setup) [27]. The measurement protocol goes as follows: by cooling the sample from above its Néel temperature in a magnetic field $B = \pm 0.4$ T (field cooling) perpendicular to the surface plane, we set the magnetic state of Cr. Subsequently, this state is read by exploiting the anomalous Hall resistance versus temperature, raised with a constant rate of 2 K/min. A close-circuit He cryostat was used to this scope.

Figure 2 reports the transport properties of a Pt(2)/Cr(50) bilayer (thickness in nm) measured with a ±20 mA probing current. Figure 2(a) shows the longitudinal resistivity $\rho_{xx}$ of the sample (empty dots) and its first derivative with respect to the temperature (black line) in the 20-330 K temperature ($T$) range. Here, it is crucial to consider that the formation of the SDW in Cr determines the opening of a gap in the Fermi surface along the direction of the wave vector [6]. This results in a reduction of the carrier density, which produces a "kink" in the resistivity versus $T$ curve. Following Ref. [15] the Néel temperature $T_N$ can be associated to the local maximum in $d\rho/dT$. From Fig. 2(a) we find $T_N$ = 290 K±5 K, which is slightly inferior to bulk Cr $T_N$ = 311 K and comparable to other epitaxial thin films [16].

The same information can be extracted from the ordinary Hall resistivity $\rho_{xy}$ since it is related to the carrier density $n$ by the simple relation coming from the Drude model $\rho_{xy} = -B/ne$ ($e$ is the electron charge), where $\rho_{xy} = R_{xy}t$ ($t$ is the film thickness). The temperature dependence of $n$ (empty squares) and its first derivative (black line) with respect to $T$ are shown in Fig. 2(b). For the reasons given above, the increase of $dn/dT$ below 290 K is consistent with the AFM phase transition of Cr [16], in agreement with the value established from longitudinal resistivity measurements.

Finally, we report the difference ($\Delta R_{xy}$) between the anomalous Hall resistances measured at remanence in states prepared by opposite field-coolings ($B=\pm 0.4$ T), shown in Fig. 2(c). Measurements at remanence permits to exclude a transverse contribution arising from ordinary Hall and to isolate the effect related to the memory of the system achieved through the manipulation of the AF state. The small signal, in the order of hundreds of nano-ohms, can be reproducibly detected and is found to be dependent only on the field-cooling direction. $\Delta R_{xy}$ decreases with temperature, approaching zero in correspondence of the temperature $T^*= 290$ K, that happens to be coincident with the previously determined $T_N$, considering the experimental error bar.

Below *T*\*, the application of magnetic fields does no longer affect the transverse resistance at remanence, thus excluding the possible influence of ferromagnetic contaminations on the signal. On the contrary, the robustness against external fields can be explained by the negligible magnetic susceptibility of Cr in the AFM phase. This strict correspondence between Cr magnetic order, established from ordinary transport measurements, and the presence of an AHE clearly relates the origin of the latter to the antiferromagnet state. We emphasize that conventional magnetometry does not allow to observe this phase transition. No change in the magnetic signal is observed with Vibrating Sample Magnetometry (VSM) across the phase transition on an unpatterned Pt(3)/Cr(75) 5x5 mm$^2$ sample. Considering the experimental error of 1 μemu, this sets the upper limit to the average magnetic moment in Cr and Pt to 0.001 μ$_B$/atom and 0.03 μ$_B$/atom, respectively. The latter is still compatible with the predicted induced moment by DFT with a mediated value smaller than 0.05 μ$_B$/atom.

To understand if the effect originates within the Pt layer, a first hint comes from the comparison of samples with different Cr and Pt thicknesses (Figure 3). Δ$R_{xy}$ at 20 K is almost identical for samples with 25, 50 and 75 nm thick Cr films and 3 nm thick Pt overlayer, with an average value of 680 ± 120 nΩ. In the good-metal regime, the anomalous Hall resistance is proportional to the magnetization of the sample and independent on the sample resistivity [13]. Being Δ$R_{xy}$ not related to Cr thickness, a bulk contribution should be therefore excluded. Instead, when a thinner Pt layer (0.6 nm, corresponding to 4 monolayers) is considered, a smaller Δ$R_{xy}$ = 275 ± 30 nΩ is found between states written with opposite field. We incidentally note that Δ$R_{xy}$ of the thinnest Cr sample (25 nm) approaches zero already below 200 K. This is consistent with the reported increase of Cr transition temperature with thickness in epitaxial thin films [15].

A second fact supporting of the interfacial origin comes from the insertion of a gold spacer between Pt and Cr. In this case, DFT calculations do not expect proximity induced moment in Au/Cr, at variance with Pt/Cr systems (See Supplemental Material S.2 for the magnetic profiles). This is in excellent agreement with experimental results: as can be seen in Fig. 3, the Pt(3)/Au(3)/Cr(50) sample does not show any detectable signal compared to any of the Pt/Cr samples. We emphasize that the same Cr(50) film was used for the comparison with Au, that was deposited on only half of the sample with the help of a mechanical shutter. Hence, the different signal cannot be ascribed to Cr quality which is confirmed to be the same by resistivity and ordinary Hall measurements (See Supplemental Material S.4 for the data). Moreover, the AHE suppression with the interposition of the diamagnetic Au interlayer allows to exclude any significant contribution to the signal given by spin-Hall magnetoresistance [28], since any phenomena related to spin-current generation in Pt would result unaffected as the spacer thickness is much thinner than the spin diffusion length in Au (λ> 30 nm [29]).

While the different signal coming from Pt/Cr and Au/Cr interfaces could be justified with the different proximity induced magnetization, the temperature trend of the anomalous Hall resistance cannot be uniquely explained as due to this net magnetic moment.

In a conventional picture the transverse resistivity in a system with magnetization *M* follows the relation [30,31]

$$\rho_{xy} = R_0 H + \alpha M \rho_{xx0} + \beta M \rho_{xx0}^2 + b M \rho_{xx}^2 \qquad (1)$$

with $\alpha, \beta$ and $b$ being the skew scattering, the side jump, and the intrinsic term, respectively, $R_0$ the ordinary Hall coefficient, $\rho_{xx0}$ the zero-temperature residual resistivity, and $\rho_{xx}$ the total longitudinal resistivity.

According to the scaling relation for $\rho_{xy}$, the temperature variation of the anomalous Hall resistivity is linked to the thermal dependence of the magnetization and of the longitudinal resistivity. It is worth pointing out that the amplitude of $\rho_{xx}$ is only doubled when moving from 50 K to about 300 K (See Fig. 2(a)). Such observation implies that it should be the magnetization to play the major role in setting the thermal dependence of the anomalous Hall resistivity. However, a closer inspection of the thermal profile of $\rho_{xy}$ indicates few distinct marks (see Fig. 3). $\rho_{xy}$ does not follow a Curie-Weiss model $\rho_{xy} \propto 1/(T - T_N)$ but, on the contrary, exhibits a smooth switching close to the antiferromagnetic transition. Moreover, the anomalous Hall resistance keeps increasing from 150 K to 50 K, in a region where the longitudinal resistivity has only minor variations and the magnetization due to the magnetic proximity should also be saturated because the sublattice antiferromagnetic magnetization has reached its maximal value.

Since uncompensated magnetization alone fails to account for these observations, an additional topological contribution has to be considered. Concerning this term, one possible scenario would point to the presence of a non-collinear antiferromagnetic pattern developing at the Pt/Cr interface. Such possibility is however ruled out as the theoretical investigation reports a strong magnetic anisotropy and the Dzyaloshinskii-Moriya coupling is small at the Cr/Pt interface [22]. For this reason, we consider the possibility of having a topological contribution due to a non-trivial Berry curvature of the bands arising from the modification of the electronic structure due to the antiferromagnetic order and the magnetic proximity.

Hence, we investigate a multiband low-energy tight-binding model which includes the bands at the Fermi level having Pt and Cr character, the SOC at the Cr site, and a magnetic interaction that takes into account the layer dependent magnetization close to the interface both in the Pt and Cr regions. Then, we determine the intrinsic geometric contribution to the anomalous Hall conductance (AHC) by evaluating the Berry curvature of the Bloch bands. The AHC is proportional to the anomalous Hall resistance experimentally measured. The analysis is performed for different spatial profiles of the magnetization at the Pt side of the heterostructure with parallel or anti-parallel magnetic moments in neighboring layers (see inset of Fig. 4). Moreover, in order to assess the role of an interface a comparison is made by replacing Pt with Au. Since the magnetization in each layer is uniform, the temperature dependence simulation of the AHC in the heterostructure can be performed by assuming a Stoner model with $M = M_0 \left[1 - \left(\frac{T}{T_N}\right)^2\right]^{1/2}$, where $M_0$ is the zero temperature magnetization. Concerning the layer dependent $M_0$ close to the interface, we take as reference the values obtained within the ab-initio analysis.

Representative results for the AHC are shown in Figure 4 where we compare three cases: i) Cr/Pt interface with a sign modulation of the magnetization in the Pt layers (red curve), ii) Cr/Pt with a uniform penetration of the magnetization in the first layers close to the interface (black curve), iii) Cr/Au interface with weak ferromagnetic proximity but higher electron density in the Au subsystem (orange curve). Due to the presence of non-trivial sources of Berry curvature, we find that an overall positive amplitude of the AHC is generally obtained assuming different types of profiles for the magnetization at the Pt/Cr interface and within the Pt subsystem, accordingly with the indications from the ab-initio analysis. On the other hand, for the case of Au/Cr interface the AHC is negative in the whole range of temperatures below the magnetic transition temperature. Although we are not aiming at achieving quantitative matching between theory and experiments, the outcomes are robust and, remarkably, due to its topological origin, the AHC is not much dependent on the character of the magnetic proximity. Furthermore, as a distinctive mark, the AHC conductance can be observed even in the case of complete compensation of the total magnetization for collinear magnetic moments both in the Pt/Cr and Au/Cr bilayers.

Recently, it was shown that in a cubic environment partially occupied $t_{2g}$ bands, in presence of strong SOC and breaking of the time symmetry, due to ferromagnetism, can also generate AHE in layered systems [32]. We observe that, although we deal with antiferromagnets, similar electronic mechanisms are at work here for the Cr/Pt heterostructure with two effective sources for the AHE: the magnetic Pt and the Cr interface layers. In bulk Cr, we have a small SOC and the topological contributions to the AHC of different antiferromagnetic layers substantially cancel each other. At the Cr/Pt interface, we have an imbalance of the Cr magnetism and the presence of a large atomic SOC. These two ingredients generate the AHE from the Cr interface layer.

On the other hand, gold is less effective in doing so because large onsite energy makes it hybridize less with interfacial Cr layers. Consistently, from our DFT calculations, gold does not show any intrinsic magnetic moment (See Supplemental Material S.2 for the magnetic profile). The thermal dependence of M the can also lead to a sign change of the AHE, depending on the details of magnetic profile of Pt, see Fig. 4.

In conclusion, DFT and electrical measurements agree on the existence of a spin-polarized Pt layer at the interface with Cr with out of plane magnetization. At the interface, Cr and Pt influence each other. In closer detail, the main effect of Cr on Pt is the induced magnetization profile. On the other hand, Pt induces a large MCA on the Cr side, favoring the out of plane component of the magnetization. Exploiting the transport properties related to SDW ordering we can demonstrate a correspondence between AHE coming from the Pt/Cr interface and AFM ordering in Cr. Finally, we calculate the Berry curvature of the spin polarized electronic structure showing the presence of the topological contribution to the AHE which can justify the occurrence and the temperature dependence of the AHE.

While the results reported in this paper highlight the specific role of the interface with Pt, we foresee similar physical scenarios with other NM metals reported to have some proximity magnetization like Pd [33], Ir,W [10], or V [34], thus making Cr a paradigmatic testbed to study these interface effects. At the same time, we expect AHE coming from interfaces to be generally exploitable in a large variety of antiferromagnetic systems, either metallic or insulating, making Hall measurements a powerful probe of antiferromagnetism both for fundamental studies and in view of applications.

**ACKNOWLEDGMENTS**

We thank R. Bertacco, X. Marti, D. Di Sante, M. Bragato, S. Achilli, and M. I. Trioni and for fruitful discussion. This work was partially performed at Polifab, the micro and nanofabrication facility of Politecnico di Milano. This work was partially funded by Fondazione Cariplo via the project Magister (Project No. 2013-0726). The work is supported by the Foundation for Polish Science through the IRA Programme co-financed by EU within SG OP. C. A. was supported by CNR-SPIN via the Seed Project CAMEO. W.B. acknowledges partial support by Narodowe Centrum Nauki (NCN, National Science Centre, Poland) Project No. 2016/23/B/ST3/00839. We acknowledge the CINECA award under the ISCRA initiative IsC43 "C-MONAMI" and IsC54 "CAMEO" Grant, for the availability of high-performance computing resources and support.

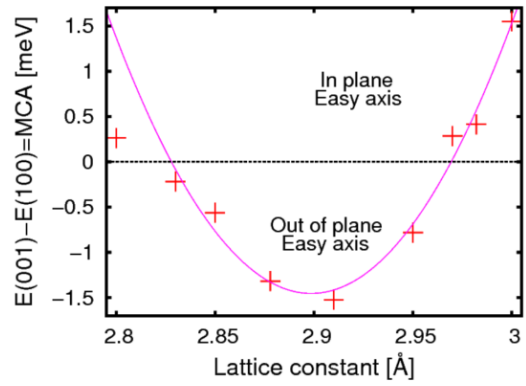

**Figure 1.** MCA of the Pt/Cr heterostructure as function of lattice constant (red points). Positive (negative) values of the MCA indicate an in plane (out of plane) easy axis, respectively. The purple line is a guide for the eyes.

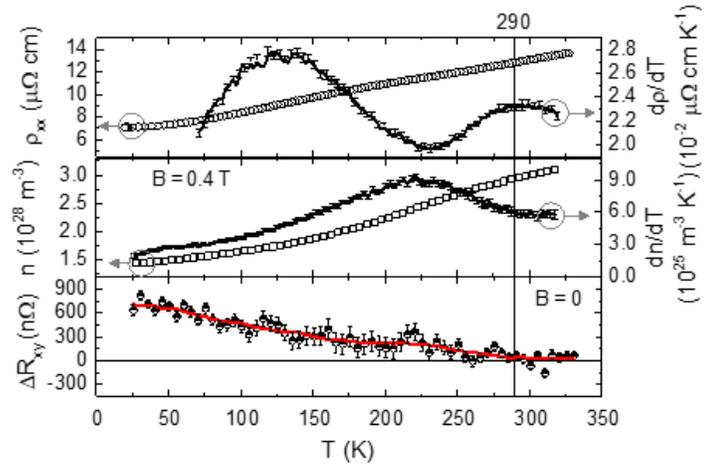

**Figure 2.** Transport properties of a Pt(2)/Cr(50) Hall cross as a function of temperature. (a) resistivity (empty dots) and first derivative (black line). (b) number of carriers from ordinary Hall measurements (empty squares) and first derivative (black line). (c) anomalous Hall resistance difference taken at two opposite amplitude of the magnetic field. The red line is a guide for the eye.

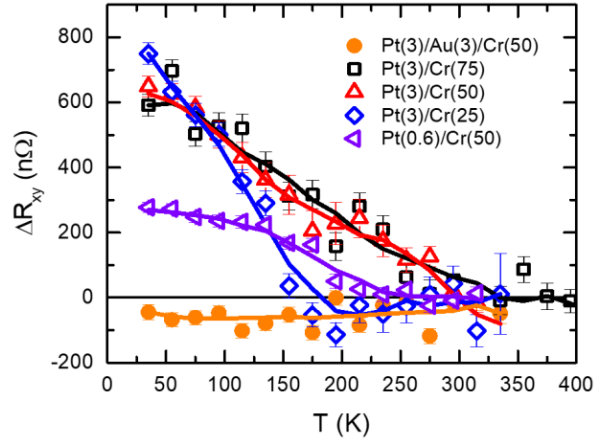

**Figure 3.** Difference of the anomalous Hall resistance taken at two opposite values of the magnetic field as a function of temperature in Pt/Cr and Au/Cr bilayers. Nominal thicknesses are expressed in nm. Error bars indicate the standard error on multiple samples. The lines are only a guide for the eye.

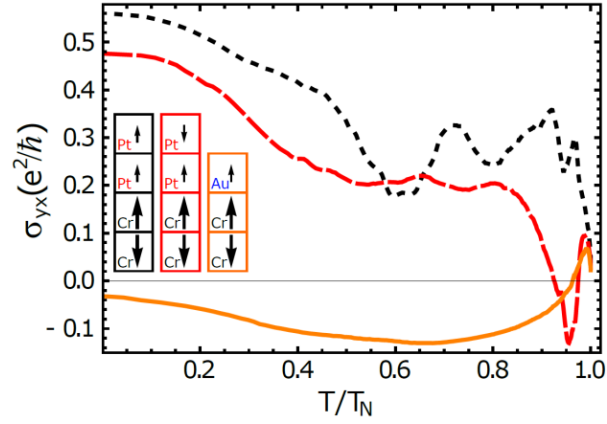

**Figure 4.** Thermal dependence of the AHC obtained by theoretically determining the Berry curvature of the d-bands for the Cr/Pt and Cr/Au layered structures assuming various magnetic profiles in the Cr, Pt and Au interface layers. The proximized magnetization for Pt or Au has a smaller amplitude than that of Cr and it can have a ferromagnetic (dotted black) or antiferromagnetic (dashed red) profile, as schematically reported in the inset. The Cr/Au interface (solid orange) corresponds to a physical configuration for the magnetization with a small leaking of magnetization in the Au interface layers.

# Supplemental material to "Anomalous Hall effect in antiferromagnetic/non-magnetic interfaces"

M. Asa,[1] C. Autieri[2,3], R. Pazzocco[1], C. Rinaldi[1], W. Brzezicki[3], A. Stroppa[4], M. Cuoco[5], S. Picozzi[2] and M. Cantoni[1]

[1] *Department of Physics, Politecnico di Milano, Via G. Colombo 81, 20133, Milano, Italy*

[2] *Consiglio Nazionale delle Ricerche CNR-SPIN, c/o Univ. "G. D'Annunzio", 66100 Chieti, Italy*

[3] *International Research Centre MagTop, Institute of Physics, Polish Academy of Sciences, Aleja Lotników 32/46, PL-02668 Warsaw, Poland*

[4] *Consiglio Nazionale delle Ricerche CNR-SPIN, c/o Univ. L'Aquila, 67100 L'Aquila, Italy*

[5] *Consiglio Nazionale delle Ricerche CNR-SPIN, c/o Universitá di Salerno- Via Giovanni Paolo II, 132 - 84084 - Fisciano (SA), Italy*

## S.1: The DOS of the Cr/Pt for different spin orientation

In order to study the surface, we add 20 Å of vacuum to settle a setup for accurate calculations. We optimize the internal degrees of freedom by minimizing the total energy to be less than $10^{-6}$ eV and the remaining forces to be less than 20 meV/Å. A 16×16×1 k-point Monkhorst-Pack grid is used for the relaxation and the calculation of the magnetocrystalline anisotropy. In addition, we use a 20×20×1 k-point for the determination of the density of states (DOS) with a smearing of 0.05 eV and an energy grid with a step of 0.1 meV. The SOC is included to calculate the magnetocrystalline anisotropy as a function of the lattice constant. We calculate the layer projected DOS for the MgO/Pt/Cr heterostructure to make an interface of the Pt with the Cr and the MgO. The plot of the difference between the DOS of the system with in plane and out of plane easy axes is reported in Fig.1.

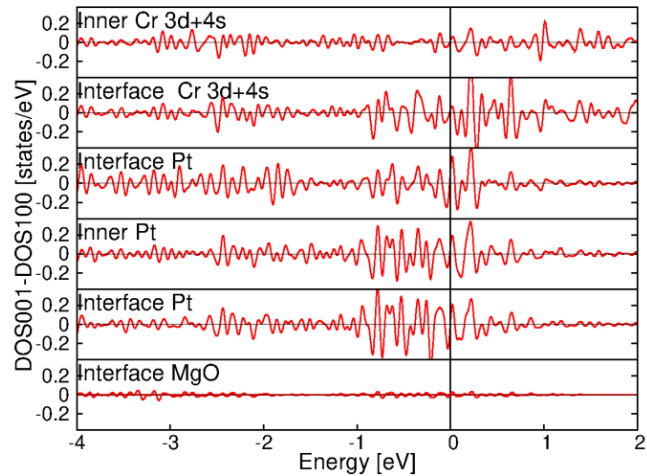

**Figure 2:** Difference between the layer projected DOS for the system calculated with the out of plane (001) and in plane (100) easy axes.

The width of the oscillation is proportional to the effect of the SOC in the different layers. Large oscillations are present in any Pt layer. The interface MgO does not show any oscillations. The Cr layers show large oscillations for the interface layer and smaller oscillations for the inner layers. Therefore, the Pt induces a large SOC on the interface layer and the contribution to the MCA mainly comes from this first layer of Cr interfaced with Pt.

### S.2: The magnetic profile for the Pt/Cr and Au/Cr interfaces

The entire profile of the Pt(N)/Cr and Au(N)/Cr magnetizations is reported in Fig. 2. The Cr magnetic moment is almost independent from the number of the 5d layers. At the surface the magnetic moment is close to 2 $\mu_B$ and the magnetization of the inner layers oscillates. The main difference in the Cr layers between the two 5d systems is the different magnetization of the interface that is larger for the Au interface.

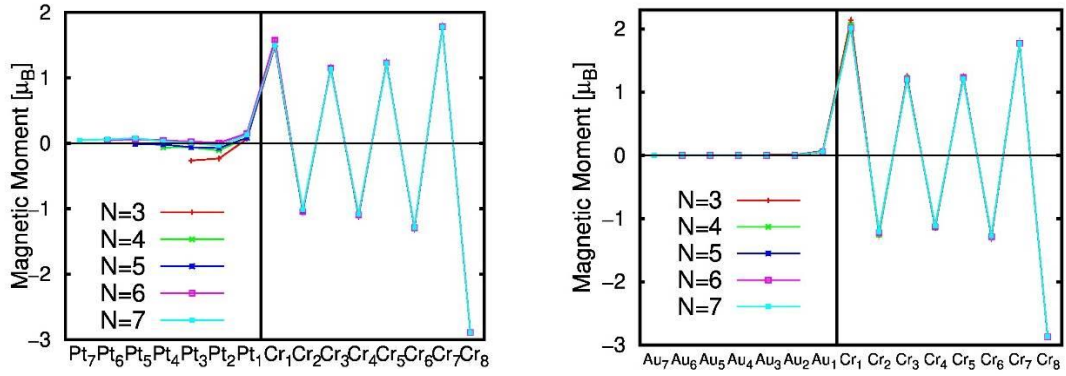

**Figure 2**: Entire magnetic profile for the Pt/Cr (a) and Au/Cr (b) heterostructure. The $Cr_1$ ($Pt_1$, $Au_1$) is the Cr (Pt, Au) interface layers. and the $Cr_i$ ($Pt_i$, $Au_i$) is the i-th layer from the interface.

A magnification of the magnetic profile of the Au/Cr heterostructure is shown in the Fig. 3. The interface gold layer presents a small magnetization of 0.06 $\mu_B$ independently on the Au thickness, while no magnetization is present for the other layers. Since this magnetization goes rapidly to zero far from the Cr layers, we propose that magnetization of the interface Au is not related to the intrinsic magnetic moment of gold but it is simply due to the tails of the magnetization of the Cr atoms. The AHE measured for the Au/Cr system can be entirely attributed to the Cr interface. This explains also the smaller value of the AHE that arises from one single Au layer, while in the case of the Pt/Cr system many Pt layers contribute to the AHE. The different sign of the AHE between Pt/Cr and Au/Cr also suggests that the main contribution to two cases is different.

A magnification of the magnetic profile of the Pt/Cr heterostructure is shown in the Fig. 4. The total magnetic moment of the Pt region is -0.429, -0.147,-0.084, 0.336 and 0.315 $\mu_B$, respectively, for N=3,4,5,6,7. The total magnetic moment of the Au region, instead, is basically always equal to the induced magnetization of 0.06 $\mu_B$.

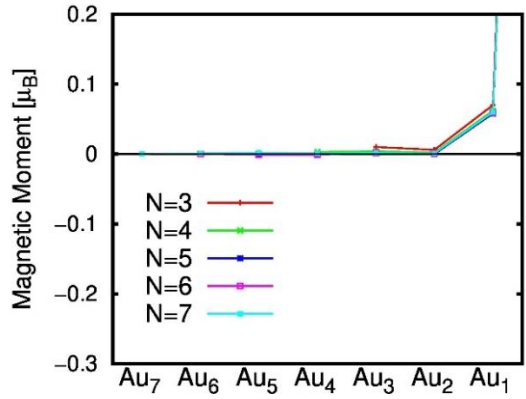

**Figure 3:** Magnetic profile of the Au(N)/Cr interface. The $Au_1$ is the Au interface layers. The $Au_i$ is the i-th layer from the interface.

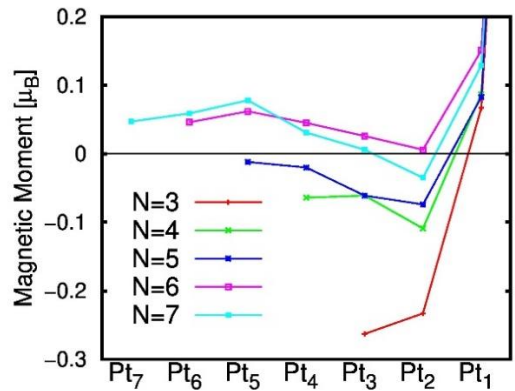

**Figure 4:** Magnetic profile of the Pt(N)/Cr interface. The $Pt_1$ is the Pt interface layers. The $Pt_i$ is the i-th layer from the interface.

### S.3: Electrical measurements in Pt/Cr

Both longitudinal and transverse resistance measurements were performed as a function of temperature using a current-source/nanovoltmeter pair Keithley 6221/2182A operating in Delta Mode. Essentially, the Delta Mode automatically triggers the current source to alternate the signal polarity, then triggers a nanovoltmeter reading at each polarity. This current reversal technique cancels out constant thermoelectric offsets, ensuring the results reflect the true value of the voltage. A relay switching matrix was used to acquire both transverse and longitudinal measurements at the same time. The two contact configurations required for geometrical offset subtraction [1] are pictured in Fig. 5, while the Van der Pauw measurements of resistivity [2] were performed using the two configurations showed in Fig. 6.

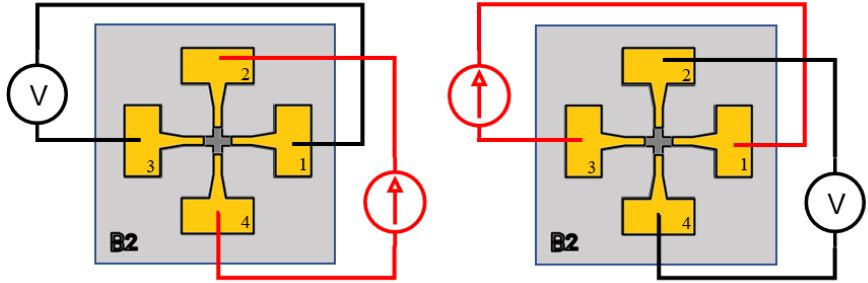

**Figure 5:** Transverse resistivity measurement configuration in Hall crosses.

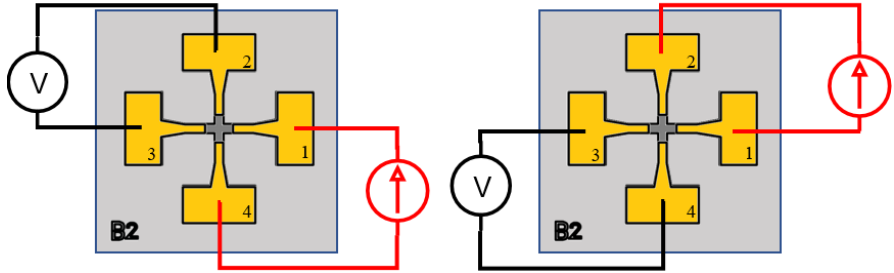

**Figure 6:** Van der Pauw configuration for the measurement of resistivity in Hall crosses.

Representative measurements of transverse resistance ($R_{xy}$ in the main text) in Pt(3)/Cr(75) are shown in Fig. 7. Here both geometrical and thermal offset compensation techniques are applied. Nevertheless, a temperature-dependent offset of non-magnetic origin, whose amplitude is often comparable to the small magnetic signal coming from the Pt/Cr interface, is observed. This uncompensated offset is different for each sample and presumably comes from minor non-uniformities in the sample. This spurious signal is fully reproducible over multiple subsequent measurements and can be subtracted as for the data reported in Fig. 2(c) and Fig. 3 of the main text.

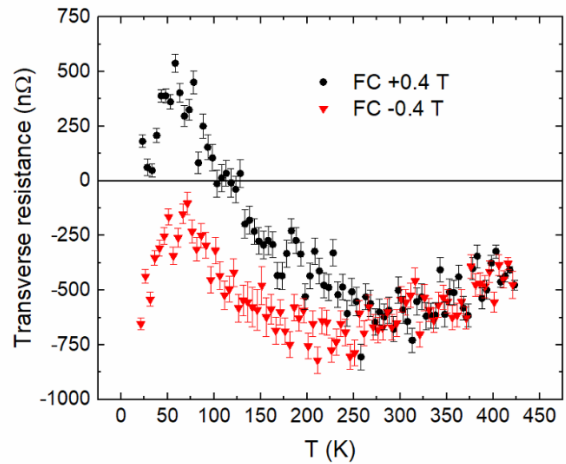

**Figure 7**: Raw measurement of transverse resistance measured in P(3nm)/Cr(75nm) following field cooling at +0.4 T (black points) and at -0.4 T (red triangles)

## S.4: Comparison of transport in Pt/Cr and Au/Cr

In Fig. 3 of the main text we compared Pt(3)/Au(3) and Pt(3) capping layers on top of the same Cr(50) thin film. To foster the conclusion that the different transverse resistance signal at remanence is related only to the different interface, we present here the transport properties of the whole heterostructure (largely dominated by the chromium layer) indicating in both cases a solid antiferromagnetic behavior below 290 K.

Figure 8 reports the longitudinal resistivity and its first derivative for Au-capped (left) and Pt-capped (right) chromium devices. In both cases the local maximum in the first derivative observed at 290 K can be used as a reference for the antiferromagnetic transition of Cr.

The same signature of the gap opening arising from the spin density wave in Cr can be observed as well, independently on the nature of the interface, in the carrier density extrapolated from Hall measurements in applied magnetic field ($B = \pm 0.4$ T) shown in Fig. 9.

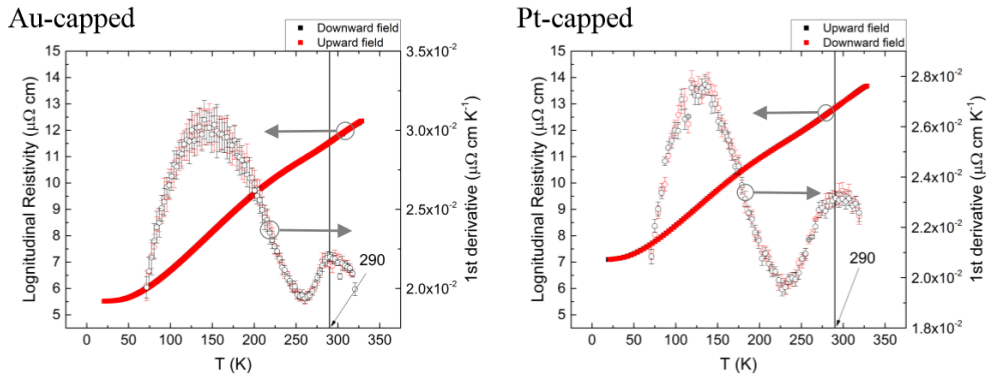

**Figure 8:** Longitudinal resistivity and its first derivative in Pt(3)/Au(3)/Cr(50) (left) and Pt(3)/Cr(50) (right).

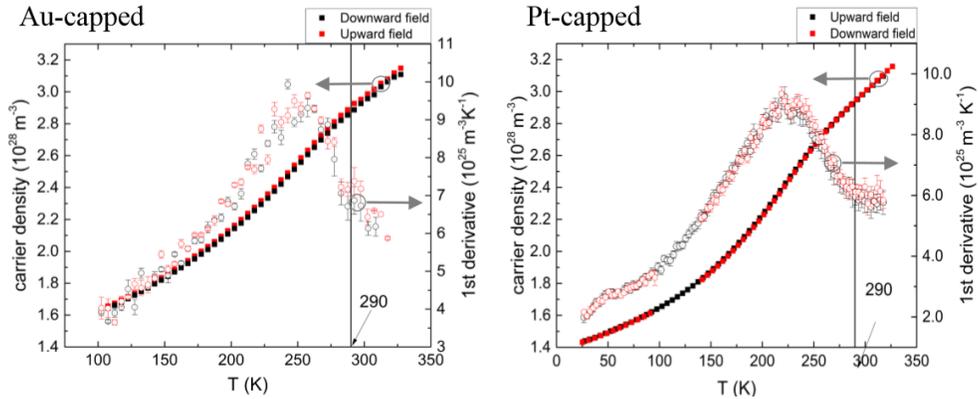

**Figure 9**: Carrier density and its first derivative in Pt(3)/Au(3)/Cr(50) (left) and Pt(3)/Cr(50) (right).

**References:**
[1] P. Daniil and E. Cohen, J. Appl. Phys. 53, 8257 (1982)
[2] L. J. van der Pauw, Philips Technical Review (1958)